\documentclass{aa}
\usepackage{epsfig}

\def\msol{M_\odot}
\def\mjup{M_{\rm J}}
\def\rjup{R_{\rm J}}

\def\te{T_{\rm eff}}

\def\lsol{L_\odot}
\def\simgr{\,\hbox{\hbox{$ > $}\kern -0.8em \lower 1.0ex\hbox{$\sim$}}\,}
\def\simle{\,\hbox{\hbox{$ < $}\kern -0.8em \lower 1.0ex\hbox{$\sim$}}\,}
\def\beq{\begin{equation}}
\def\eeq{\end{equation}}

\begin{document}


\title{Evolutionary models for cool brown dwarfs and
extrasolar giant planets. 
The case of HD 209458}
 \author{I. Baraffe\inst{1,2}, G. Chabrier\inst{1}, T. S. Barman\inst{3},
F. Allard\inst{1}
\and 
P.H. Hauschildt\inst{4}
}

\offprints{I. Baraffe}

\institute{C.R.A.L (UMR 5574 CNRS),
 Ecole Normale Sup\'erieure, 69364 Lyon
Cedex 07, France (ibaraffe, chabrier, fallard@ens-lyon.fr)
\and Max-Planck Institut f\"ur Astrophysik, Karl-Schwarzschildstr.1,
D-85748 Garching, Germany
\and
Department of Physics, Wichita State University, Wichita, KS
67260-0032 (travis.barman@wichita.edu)
\and
Hamburger Sternwarte, Gojenbergsweg 112,
21029 Hamburg, Germany (phauschildt@hs.uni-hamburg.de)}

\date{Received /Accepted}

\titlerunning{Evolutionary models for cool brown dwarfs}
\authorrunning{Baraffe et al.}
\abstract{We present evolutionary models for cool brown dwarfs
and extra-solar giant planets. The models reproduce the main trends
of observed methane dwarfs in near-IR color-magnitude diagrams. 
We also present evolutionary models for  irradiated planets,
coupling for the first time irradiated atmosphere profiles and inner
structures. We focus on HD 209458-like systems and show that
irradiation effects can substantially affect the radius of
sub-jovian mass giant planets.
Irradiation effects, however, cannot alone explain  the large observed radius
of HD 209458b. Adopting assumptions which optimise irradiation effects 
and taking into account the extension  of the outer atmospheric
layers, we still find $\sim$ 20\% discrepancy between observed
and theoretical radii. 
An extra source of energy seems to be required 
to explain the observed value of the first transit planet.
}

\maketitle

\section{Introduction}

The past decade was marked by two major discoveries in the field of
stellar and planetary physics:
the detections of the first unambiguous brown dwarf (BD) GL 229B (Oppenheimer et al . 1995) and the first extrasolar giant planet (EGP)
51 Peg b  (Mayor \& Queloz 1995). The near-IR spectrum of GL 229B was
found to be dominated by strong methane absorption bands, looking
more similar to Jupiter than to late type-stars. 
 On the other hand, the surprisingly small orbital separation between
 51 Peg b and 
its parent star suggests that the planet should be affected
by irradiation
and that, given the expected large surface temperature, its atmospheric
properties should resemble more the ones of relatively hot brown dwarfs than
the ones of jovian planets.

Since then, about thirty
methane dwarfs (or the so-called T-dwarfs) have been identified, due mainly
to the near-IR surveys 2MASS (Burgasser et al. 1999), SDSS (Strauss et
al. 1999) and the VLT (Cuby et al. 1999). The radial velocity
technique has 
now revealed more than 100 EGPs in orbit around nearby stars 
(see Hubbard, Burrows,
Lunine, 2002 for a review and references therein), with a large fraction
($\sim$ 10\%-20\%) being extremely close (less than 0.06 AU) 
to their parent star. 
The mass of substellar companions detected by radial velocimetry 
extends well above the deuterium burning minimum mass 
$0.012 \msol = 12 \mjup$ (Saumon et al., 1996; Chabrier et al., 2000a). This mass is often used as
the boundary between  planets and brown dwarfs,
more for semantical than physical reasons.
On the other hand isolated objects with planetary masses are
now  discovered in young stellar clusters, down to a few $\mjup$,
as recently reported by Zapatero et al. (2002) in $\sigma$ Orionis.
These observations suggest that there is an overlap between the mass range
of the least massive brown dwarfs and of the most massive giant planets.
In principle, different formation processes should distinguish planets
from brown dwarfs. However,  such a distinction is difficult to
characterize in terms of atmospheric, structural and cooling
properties since both
types of objects have convective interiors with 
essentially a metallic H/He mixture. The signature of a central rock+ice core, like in solar giant
planets, would be the clear identification of a planet. 
 The presence of a core can  affect the radius of a planet, yielding
a smaller planetary radius than in the absence of a core. For 1 $\mjup$, the
effect is about 5\% on the radius for a core mass $<$ 0.06 $\mjup$
(see Saumon et al. 1996). The presence of this core  can be 
 inferred  from the accurate characterization 
of the gravitational moments of the object,
and such an observation is currently not feasible for EGPs. 
In addition, both giant planets and brown dwarfs have
atmospheres dominated by molecular absorption and the
effects of cloud formation. 
Although frustrating from an observational point of view, these
similarities imply that the general cooling theory for BDs,
involving detailed models of the atmosphere and inner structures, can
be applied to EGPs.
In terms of cooling properties, this general theory can even be applied to Jupiter, 
as emphasized by Hubbard et al. (2002). Additional observational constraints,
as provided by spacecraft encounters or
by  direct probes (e.g Galileo), 
have lead to refinements of the models (heavy element core, 
non-standard chemical composition).
As mentioned above, such constraints are, unfortunately, far from being accessible for EGPs.

Much effort has been devoted to the modeling of substellar objects 
 during the past decade, improving our understanding 
of cool atmospheres (see Allard et al. 1997 for a review),
of the role of dust   (Tsuji et al. 1999; Burrows et
al. 2000; Ackerman \& Marley 2001, Allard et al. 2001;
Marley et al. 2002), of irradiation  (Saumon
et al. 1996; Seager \& Sasselov 1998; Sudarsky et al. 2000; 
Barman et al. 2001), and of their inner structure and evolutionary
properties (Burrows et al. 1997 ; Chabrier et al. 2000b; see
Chabrier \& Baraffe 2000 for a review). One remaining major
challenge in the theory is the description of dynamical processes
of grain formation and diffusion necessary to understand
the transition between L-dwarfs and T-dwarfs,  which is expected to take
place at $\te \sim$ 1300K - 1700K. 
The former objects are better reproduced by
dusty atmosphere models, whereas the later  are better reproduced
by dust-free (or partly dusty) models. The recent observations of 
L/T dwarfs at the transition clearly indicate that complex processes 
take place in the atmosphere of these objects (see e.g Burgasser et al. 2002).
Another important challenge is the modeling of irradiation effects,
which are expected to affect  the spectra of close-in
EGPs, and may also affect their inner structure and cooling properties. 
The recent discovery of the planet HD209458b transiting its parent
star (Charbonneau et al. 2000) provides a unique test to explore 
such effects, 
since its mass and radius can be determined with high accuracy
from the modeling of the transit lightcurve. According to the most
recent determination (Cody \& Sasselov 2002), the mass and radius
of the planet are estimated to be $m = 0.69 \pm 0.02 \mjup$ and
$R=1.42^{+0.10}_{-0.13} R_{\rm J}$. 

Evolutionary models including crude estimates of the effects of irradiation 
 on planet atmospheres suggest that extrinsic heating is sufficient
 to maintain a larger planetary radius compared to an isolated
planet.
It has thus been argued that irradiation could explain the large
radius of HD209458b (Guillot et al. 1996; Burrows et al. 2000).
More recently, Guillot \& Showman (2002) questioned such results
and argue that the radius of HD209458b can only be reproduced if the deep
atmosphere is much hotter than what can be expected
from irradiation effects. 
However,  none of these calculations includes a {\it consistent treatment
between the irradiated atmospheric structure and the interior structure}
of the planet. Such a consistent treatment is mandatory to get reliable results
since the deep interior entropy profile, which determines the heat content of the
planet to be radiated away while it cools, is affected by the modification of the
atmospheric temperature profile due to the incoming external heat flux.
 The main goal of the present paper is to present the first such consistent calculations.
As mentioned above, in the case of non-irradiation, these calculations apply
to the evolution of cool (dust-free like) brown dwarfs, i.e. T-dwarfs, and extrasolar
giant planets far enough from the parent star for the irradiation effects on the
thermal structure of the planet to be negligible. 
This is the case of the solar
giant planets, the cooling of which is simply characterized by 
the cooling properties
of the "isolated" planet plus the heating contribution from the Sun $4\pi \sigma R_p^2T_\odot^4$,
where $R_p$ is the radius of the planet and $T_\odot$ represents the
equivalent black body temperature of the converted solar radiation
(Hubbard, 1977; Guillot et al., 1995).
In section 2 we briefly present the input physics of
non-irradiated models, describing
methane dwarfs and isolated EGPs. Apart from the impinging stellar
flux, the same input physics are
used to analyse the effects of irradiation. 
The effects on the radius and cooling properties
of giant planets are described in section 3 and results are compared to
the observed properties of HD209458b. Discussion follows in \S 4. 

\section{Non-irradiated models}

\subsection{Model description}

The main input physics involved in the present models are the same as
described in our previous works (Chabrier \& Baraffe 1997; Baraffe et
al. 1998; Chabrier et al. 2000b). The models are based on the coupling
between interior and non-grey atmosphere structures.
The treatment of dust in the atmosphere is described in detail in
Allard et al. (2001), with two limiting cases of dust treatment. The first case, referred to as
``DUSTY'', takes into account the formation of dust in the equation
of state, and its scattering and absorption in
the radiative transfer equation. Such models 
 assume that dust species remain where they form,
according to  the chemical equilibrium conditions. 
The second case, referred to as ``COND'',
neglects dust opacity in the radiative transfer equation. 
In a previous paper (Chabrier et al. 2000b), we presented the
evolutionary models based on DUSTY atmosphere models, aimed at describing
the evolution and the photometric and spectroscopic properties of early L-dwarfs. The present
paper is devoted to evolutionary models based on the COND approach,
which 
are more appropriate to objects with effective temperatures $\te \simle
1300$K, such as methane dwarfs or EGPs at large orbital separation.
These models apply when
all grains have gravitationally settled below the photosphere. A forthcoming paper will be devoted
to models taking into account 
characteristic diffusion timescales of different processes affecting
the dust stratification (e.g coagulation, gravitational settling,
convection).  These models aim at describing in particular
the transition objects
between late L-dwarfs and early T-dwarfs (see Allard 2002). 

\subsection{Evolution of cool brown dwarfs}

\begin{figure}
\psfig{file=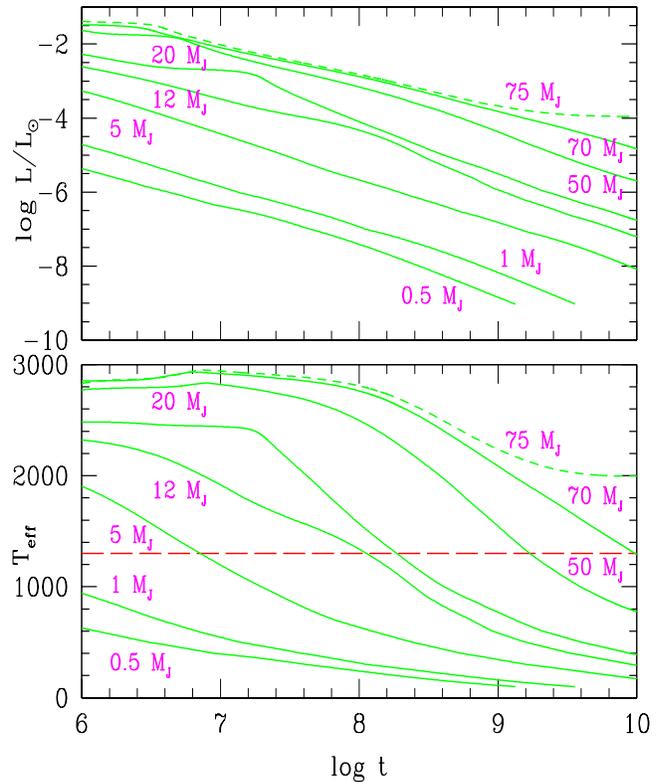,height=110mm,width=88mm} 
\caption{Evolution of the luminosity $L$ and effective temperature
$\te$ as a function of time (in yr) for different masses. Substellar
objects are indicated by solid lines and stars by short-dashed lines.
The horizontal long-dashed line indicates the limit ($\te \sim$1300K)
below which the COND
models are appropriate for the photometric and spectroscopic
description of T-dwarfs and EGPs (see text).
}
\label{fig0}
\end{figure}

A preliminary version of the COND models was  presented in
Chabrier et al. (2000b) down to 0.01 $\msol$. In the present paper,
we  extend our calculations to
$\te$ = 100 K and $m=$ 0.5 $\mjup$. The evolution
of $L$ and $\te$ as a function of time for different masses is
displayed in Fig. \ref{fig0}. The properties of the COND models for
different ages are given in Tables 1-5. 
As already stressed in Chabrier et al. (2000b), the treatment of dust
in the atmospheric models barely affects the evolutionary tracks in terms of $L$ and $\te$ as a function of time for a given mass (see Fig. 2 of Chabrier et al. 2000b). 
Consequently, although the COND models are more appropriate 
to describe the {\it spectral and photometrical properties} of substellar objects with $\te \simle$ 1300K, 
they provide a good description of their {\it cooling properties}
even at higher $\te$. In other words, it is not necessary to perform 
evolutionary calculations  with the DUSTY models above $\te \sim$ 1300K and switch to the COND models below.
An object characteristic of the present models
was recently discovered
by Zapatero et al. (2002) in $\sigma$ Orionis: S Ori 70. From a comparison
of its observed spectrum with COND synthetic spectra (Allard et al. 2001),
Zapatero et al. (2002) estimate an effective temperature  $\te \sim$ 700-1000K. If
the membership of S Ori 70 to $\sigma$ Orionis is confirmed, implying an age
$<$ 10 Myr, its mass mass should be $\simle 5 \mjup$ (see
Fig. \ref{fig0}). 

\begin{figure}
\psfig{file=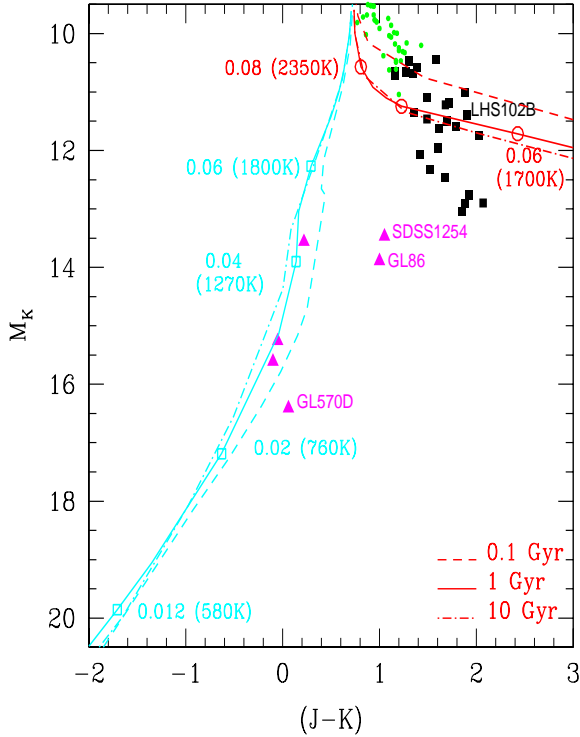,height=110mm,width=88mm} 
\caption{Color - Magnitude diagram $(J-K)$ - $M_{\rm K}$.
Observations are taken from Leggett (1992) (mostly for
M-dwarfs) and Dahn et al. (2002).
 Also shown: LHS 102B (Goldman et al. 1999), GL86 (Els et al. 2001). 
M-dwarfs are shown by dots,  L-dwarfs by filled squares and T-dwarfs by
triangles.
DUSTY isochrones (Chabrier et al. 2000b) are
displayed in the upper right part of the figure, for different ages, 
as indicated. The COND isochrones
are displayed in the left part of the figure.
Some masses (in $\msol$) and their corresponding $\te$ are indicated on the 1 Gyr
isochrones by open squares (COND) and open circles (DUSTY). 
The names of two L/T transition objects and of the faintest T-dwarf known
with parallax are indicated.
}
\label{fig1}
\end{figure}

\begin{figure}
\psfig{file=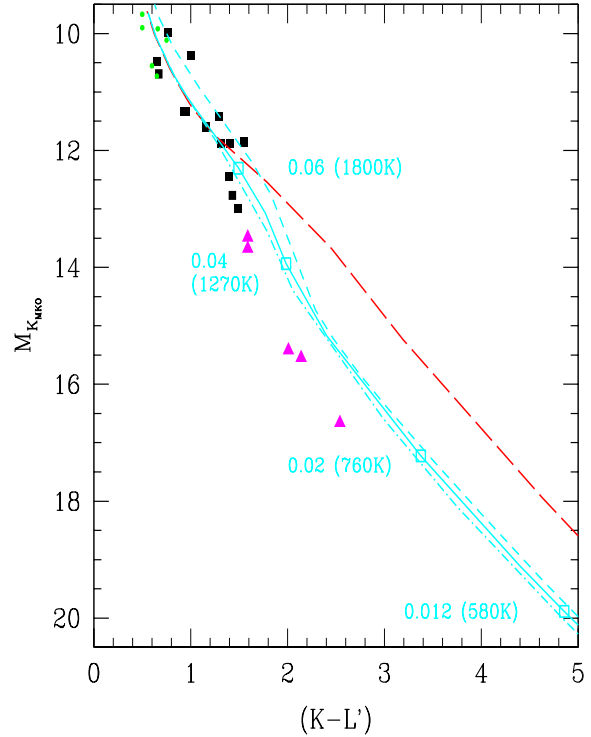,height=110mm,width=88mm} 
\caption{Color - Magnitude diagram $(K-L')$ - $M_{\rm K}$.
K is in the MKO-NIR system. 
Observations  are from Leggett et al. (2002a). Symbols are the
same as in Fig. \ref{fig1}. 
A DUSTY isochrone of 1 Gyr (Chabrier et al. 2000b) is
indicated by the long-dashed line. The COND isochrones
are displayed for 0.1 Gyr (dash), 1 Gyr (solid) and 10 Gyr (dash-dot).
Some masses (in $\msol$) and their corresponding $\te$ are indicated on the 1 Gyr COND isochrone by open squares.
}
\label{fig2}
\end{figure}

\begin{figure}
\psfig{file=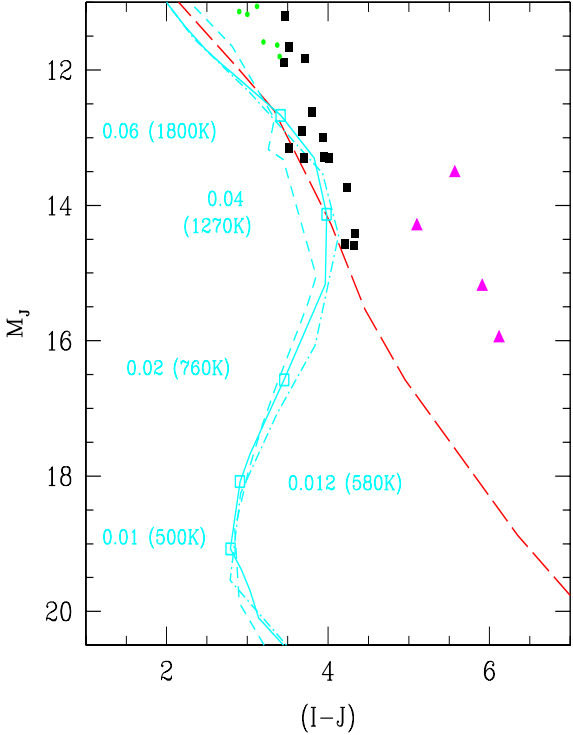,height=110mm,width=88mm} 
\caption{Color - Magnitude diagram $(I-J)$ - $M_{\rm J}$.
Observations  are  from Leggett (1992) and Dahn et al. (2002).
Symbols and curves are the same as in Fig. \ref{fig2}. 
}
\label{fig3}
\end{figure}

Several methane dwarfs have been discovered in the solar neighbourhood,
implying older ages and thus larger masses than the extreme case of 
S Ori 70. At an age of $10^8$ yr, only  objects
with masses below the deuterium burning minimum mass ($m \le 0.012 \msol$)
have $\te \simle 1300K$, whereas at 5 Gyr, it is the case
for all substellar objects with $m \simle 0.06 \msol$ (see
Fig. \ref{fig0}
and Tables 1 and 4).
Photometric observations and parallax determinations of several L- and T-dwarfs
(Els et al. 2001; Leggett et al. 2002a; Dahn et al. 2002)
now allow  a comparison with models in observational color-magnitude
diagrams (CMD),
providing stringent constraints on theoretical models (see
Figs. \ref{fig1}-\ref{fig3}).

As already noticed in Allard et al. (1996) for GL 229B, models free of atmospheric dust clouds better reproduce the near-IR photometric and spectral properties of methane dwarfs. This is illustrated in $(J-K)$ and $(K-L')$
colors in  Fig. \ref{fig1} and Fig. \ref{fig2} respectively, where
 the COND models reproduce the main observed trends.  
In Fig. \ref{fig1}, we note the two transition objects, intermediate
between L- and T- dwarfs, with  $(J-K) \sim 1$ (GL 86B: Els et al. 1999;
SDSS 1254-01: Leggett et al. 2002a, Dahn et al. 2002) and
the faintest L-dwarfs (Dahn et al. 2002), which  are not described by either
the DUSTY or COND limiting cases, and require a 
detailed treatment
of dust diffusion in the atmosphere, as mentioned in \S 2.1.

The predictions of the COND models provide a general good agreement
with observed near-IR photometry and spectra at wavelength $> 1 \mu$m
(Leggett et al. 2002b; Zapatero et al. 2002). 
The models show however shortcomings at shorter wavelength, with
a flux excess around 0.8-0.9 $\mu$m, characteristic of the I-bandpass. 
This problem is illustrated in Fig. \ref{fig3} in a $(I-J)$ - $M_{\rm J}$ CMD,
where the COND models predict significantly bluer $(I-J)$ colors
than observations. As mentioned in Allard et al. (2001),
uncertainties in the current treatment of the
far wings of the absorption lines of alkali elements (Na, K) 
at such pressures
may be responsible for this discrepancy.
No theory, however, exists to date for an accurate description of broadening of atomic
lines by collisions with H$_2$ and He. Attempts to improve current
treatments are under progress (Burrows \& Volobuyev 2002).

The correct trend of colors and spectral properties
predicted by the present models
at wavelength $> 1 \mu$m, where most of the flux is emitted
for the concerned range of $\te$,
comfort us however with their reliability
to describe extremely cool objects. 

\section{Irradiated models}

\subsection{Effect on atmosphere structure}

As mentioned in the introduction, a non-negligible fraction of EGPs orbit close to their parent star
and their thermal and mechanical structure is affected by irradiation effects.
 Therefore, a general theory of
cool substellar objects must take these effects into account. 
Recently Barman et al. (2001) have modeled irradiated atmospheres by
including the impinging radiation field in the solution of the radiative transfer equation.
As shown by these authors, for a given intrinsic luminosity,
non-irradiated planets have very different temperature structures than
irradiated planets.  Thus, substituting non-irradiated atmospheric
structures with $\te$ = $T_{eq}$ (see definition below, Eq.(8)) for
irradiated structures, as done up to now in the literature,
yields  incorrect inner boundary conditions for
evolutionary calculations (see e.g. figure 13 of Barman et al. (2001)).
Given the present lack of an accurate
treatment of atmospheric dust diffusion, the calculations 
were performed only for the  DUSTY and COND limit cases, respectively. 
The results emphasize the strong dependence of the emergent spectrum
and atmospheric structure on the presence or absence of dust.
In the absence of dust, the impinging flux can penetrate in 
deeper layers of the planet atmosphere, affecting more drastically
the inner structure of the planet
than in the dusty case. 

Except for a possible detection of sodium absorption in the atmosphere
of HD 209458b (Charbonneau et al. 2002), no constraints on the atmospheric
composition of EGPs are
available at the present time.
The only strong observational constraint available
for irradiated models  is the transit planet HD 209458b. The determination
of its mass and radius provides a stringent test to irradiated atmosphere 
 calculations and to the resulting structure and
evolution.  We thus apply our calculations of irradiated EGPs to HD 209458-like systems.

We have computed  
 a grid of irradiated atmosphere models based on the COND input physics 
described in \S 2, as in Barman et al. (2001). 
Although more appropriate for EGPs with $\te \simle 1300$ K, the COND models maximise the effect of irradiation on the inner atmosphere structure and thus
on the evolution of EPGs (Barman et al. 2001).
The grid
 covers a wide range 
of $\te$ from 40K to 100K, in steps of 20K, and from 100K to 2800K, 
in steps of 100K. It  covers a range of  
surface gravities from $\log g=2.5$ to $\log g=4.5$, in steps of 0.5 dex.
We adopt the characteristics of HD 209458,
assuming for the primary
an effective temperature $\te$$_\star$ = 6000 K,
 a radius $R_\star$ = 1.18 $R_\odot$ (Mazeh
et al. 2000; Cody \& Sasselov 2002) and an orbital separation $a$ = 0.046 AU 
(Charbonneau et al. 2000). 
As in Barman et al. (2001), we make the simplifying assumptions
that the impinging radiation field is isotropic and  the incident
flux $F_{\rm inc}$ is redistributed only over the dayside, i.e.
\begin{equation}
F_{\rm inc} = {1 \over 2} \, ({R_\star \over a})^2 F_\star, 
\end{equation}
where $F_\star$ is the total flux from
the primary (see discussion in \S 4).

\begin{figure}
\psfig{file=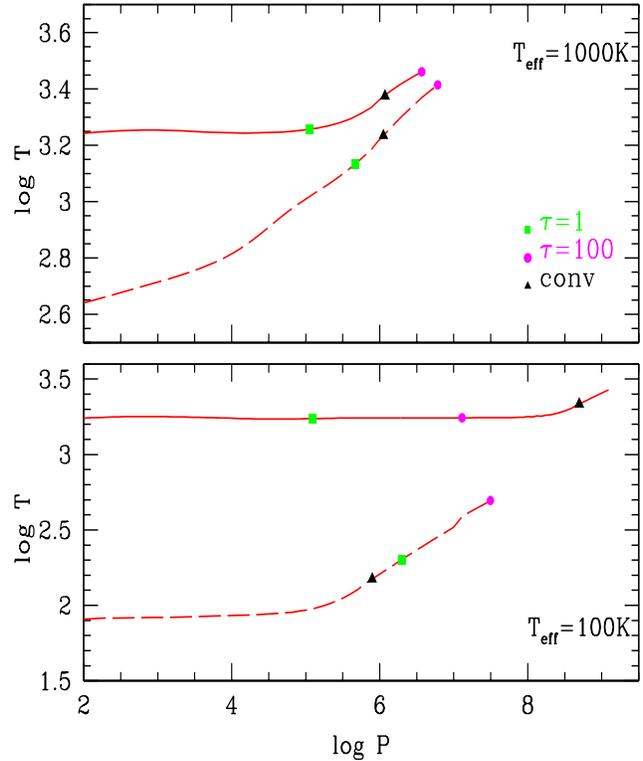,height=110mm,width=88mm} 
\caption{Effect of irradiation on atmosphere profiles, 
T (K) versus P (dyn/cm$^2$), characterized by
a surface gravity $\log g = 3.0$, $\te$ = 1000K (upper panel) and 
$\te$ = 100K (lower panel). Dashed lines correspond
to  nonirradiated structures. Solid lines are irradiated models
at a separation $a$ = 0.046 AU from a primary with $T_{{eff}_\star}$ =
6000K. The corresponding equilibrium temperature is $T_{\rm eq} \sim
1630K$.
The squares on the curve refer to optical
depth, defined at $\lambda = 1.2 \mu$m,  $\tau_{\rm std}$ = 1 and the
circles to $\tau_{\rm std}$ =100. The triangles indicate the top of
the convective zone. 
}
\label{fig4}
\end{figure}

Before proceeding any further,
we briefly re-specify  definitions
of fluxes (see e.g. Brett \& Smith 1993),
since use of various terminologies  leads to confusion.
In all cases,
the integrated {\it net flux} $F_{\rm net}$, obtained from the solution of the transfer equation,
 is the intensity integrated over
both in-coming and out-going directions $\mu$ ($\mu = \cos\theta$, where
$\theta$ is the angle of incidence). 
Assuming there is no extra source
or sink of energy (e.g no horizontal energy transfer), energy
conservation implies that all the incident energy 
coming in must go out. 
Therefore, in the case of irradiation, the in-coming flux from the parent star 
cancels out the extra out-going, absorbed and reradiated flux
due to the heating of the upper layers of
the planet atmosphere (see Fig. \ref{fig4}). 
The in-coming flux at the surface is $F_{\rm in}$ = - $F_{\rm inc}$ and
the out-going flux at the surface is 
$F_{\rm out}$ = $F_{\rm inc}$ + $\sigma \te^4$, where $\sigma \te^4$ defines the
intrinsic, unperturbed flux $\sigma \te^4$ of the initial, non-irradiated atmosphere structure.
Energy conservation thus implies:

\begin{equation}
F_{\rm net} = F_{\rm out} + F_{\rm in}=\sigma \te^4
\end{equation}

\noindent The non-irradiation case ($F_{\rm inc}=0$) corresponds to
the usual condition  $F_{\rm net} = F_{\rm out} =\sigma \te^4$.

Our atmosphere models, irradiated or not, are thus characterized by the 
parameters $\te$ and $g$. Of course, the same net flux $F_{\rm net}$
corresponds to two different atmospheric structures,
in the non-irradiated and irradiated case, because of the extra energy source
$F_{\rm in}\ne 0$ in the latter case (see Fig. \ref{fig4}).
Given the above definitions, the net flux characterizes
the {\it intrinsic luminosity}, i.e.  the rate of energy released by the planet as it contracts and cools down:

\begin{equation}
L_{int}=4 \pi R_{\rm p}^2 \sigma \te^4 = \int -T {dS \over dt} dm. 
\end{equation}

This quantity determines the cooling properties of
the planet for a given set of outer boundary conditions provided
by the atmospheric profile (see \S 3.2 below).
We stress that, in the case of irradiation, $\te$ does {\it not} characterize
the {\it total flux} emitted by the planet, which is given by:
\begin{equation}
F_{\rm out}=\sigma \te^4 \,+\, F_{\rm inc}=\sigma \te^4 \,+\, {1 \over
2}\, ({R_\star \over a})^2 F_\star
\end{equation}
\noindent Note that $F_{\rm out}$ is the important quantity for observers,
since it characterises the total radiation of
the planet, including both thermal and reflected parts of the flux.
However, we do not focus on this quantity, since
a forthcoming paper will be devoted to  spectral
properties of irradiated planets (Barman et al. 2003, in preparation).

For the sake of comparison
 with non-irradiated atmosphere profiles (see Barman et al. 2001),
we also define the quantity 
$T_{\rm therm}$ which
characterises the thermal flux $\sigma T_{\rm therm}^4$
emitted by the irradiated fraction of the planet (in the present case, the day side only). 
This quantity reads:

\begin{equation}
 \sigma T_{\rm therm}^4 = \sigma T_{\rm eff}^4 + (1-A)F_{\rm inc},
\end{equation}

\noindent where $A$ is the Bond albedo. According to the definitions
above:
\begin{equation}
F_{\rm out} = \sigma T_{\rm therm}^4 + F_{\rm refl},
\end{equation}
\noindent where 

\begin{equation}
F_{\rm refl}=AF_{\rm inc}
\end{equation}

\noindent is the reflected part of the incident flux.

Within the conditions of the present calculations (Eq(1), $a=0.046$, $\te$$_\star$ = 6000 K,
$R_\star$ = 1.18 $R_\odot$), our Bond albedo is close to 0.1
for the coolest models ($\te \sim 100$K). 
A final quantity, often used
in the literature, is
the equilibrium temperature, $T_{\rm eq}$, 
which characterizes the planet's luminosity
after having exhausted all its internal heat content (see e.g Guillot et al. 1996; Saumon et al. 1996): 

\begin{equation}
\begin{array}{l}
 T_{\rm eq}^4 =  {(1-A)\over \sigma}F_{\rm inc} \, = {(1-A)\over 2 \sigma} \, ({R_\star \over a})^2 F_\star\\
\hspace{1cm}\\
\hspace{0.8cm} \rightarrow T_{\rm therm}^4 \,\,{\rm when}\, T_{\rm eff}\rightarrow 0.
\end{array}
\end{equation}

 Note that given our definition of $F_{\rm inc}$, $T_{\rm eq}^4$ defined by 
Eq. (8) differs by a factor
2 from the definition usually used in the literature, because of the redistribution only over the day side.
Note also that $T_{\rm eq}$ and $T_{\rm therm}$ differ significantly at young ages, when the intrinsic flux of the planet is not negligible. 

The effect of irradiation on atmosphere structures is illustrated
in Fig. \ref{fig4} for different  values of the effective temperature $\te$.
As already stressed in Barman et al. (2001),
an irradiated structure characterised by $T_{\rm therm}$ can
differ  significantly from  a non-irradiated structure at the same effective
temperature $\te$ = $T_{\rm therm}$. 
This point (see also Seager \& Sasselov 1998;
Guillot \& Showman 2002) emphasizes the fact that adopting 
outer boundary conditions, for
evolutionary calculations,
from atmospheric profiles of nonirradiated
models with  $\te$ = $T_{\rm therm}$, or  $\te$ = $T_{\rm eq}$ (as e.g. 
Burrows et al. (2000)), is incorrect and yields erroneous evolutionary properties
for irradiated objects.
 
\subsection{Effect on evolution}

The main effect of irradiation on convective atmospheres and
its consequences on evolution is 
well known (see Hubbard 1977; Brett \& Smith 1993; Guillot et al 1996;
Hubbard et al. 2002). 
 The heating of the outer layers by the incident flux reduces 
the temperature gradient between these layers and the interior.
They become radiative and the top of the convective zone
is displaced  to larger depths compared to the non-irradiated case,
as clearly illustrated in Fig. \ref{fig4}. 
The inner atmosphere structure is 
hotter at a given pressure than the nonirradiated atmosphere 
model of same $\te$ (see  Fig. \ref{fig4}). In order
to match the same inner entropy, or the same values of $P$ and $T$,
characteristic of the boundary layer between the interior structure 
and the irradiated atmosphere structure, characterized by
a given $\te$ and $\log \, g$, one would need a nonirradiated
atmosphere model with
higher $\te$, i.e a larger heat loss.   
Therefore, for a given planet heat content, i.e. internal entropy, the
heat loss is reduced in the case of irradiation and the
planet maintains  a higher entropy for a longer time. 
Since for a given mass, the interior (P,T) profile and thus the entropy fix the radius,
the irradiated planet has a larger radius than the nonirradiated
counterpart at a given time, starting from the same initial configuration.
In other terms, 
gravitational  contraction, which is the dominant source of energy
of the planet, proceeds more slowly with irradiation than
without it.  

Our calculations proceed as for our low-mass star or brown dwarf
calculations, by coupling the interior and atmosphere profile at a
deep enough optical depth, which defines unequivocally the fundamental
properties of the object, $m,R,\te,L$ along its evolution $t$ (Chabrier \& Baraffe, 1997). 
The boundary condition between inner and atmosphere
structure is fixed at $\tau_{\rm std}=100$,
which corresponds to a range of pressure $P$ = 0.1 - 200 bar for the whole
range of atmosphere models \footnote{$\tau_{\rm std}$ is
defined at $\lambda=1.2 \mu m$}. 
The irradiated atmosphere models are integrated down to an
optical depth $\tau_{\rm std}=100$ for $\te \ge 1000K$ and $\tau_{\rm std}=10^5$ for $\te <$
1000K. In both cases, this is deep enough to reach the top
of the convective zone and to provide a good spatial resolution 
of these layers, even for the coolest models (see Fig. \ref{fig4}). 
In any case, the incident flux $F_{\rm inc}$ drops to zero
at $\tau_{\rm std} <<50$, well above the deepest layers of the atmosphere models.
Note that for the coolest atmospheric structures, convection does not reach the
layers corresponding to $\tau_{\rm std}=100$ (see Fig. \ref{fig4}).
 In that case, the radiative gradient
in the interior is calculated with the Rosseland means of the same atmospheric opacities, for a consistent treatment between
the interior and atmosphere thermal structures.

\begin{figure}
\psfig{file=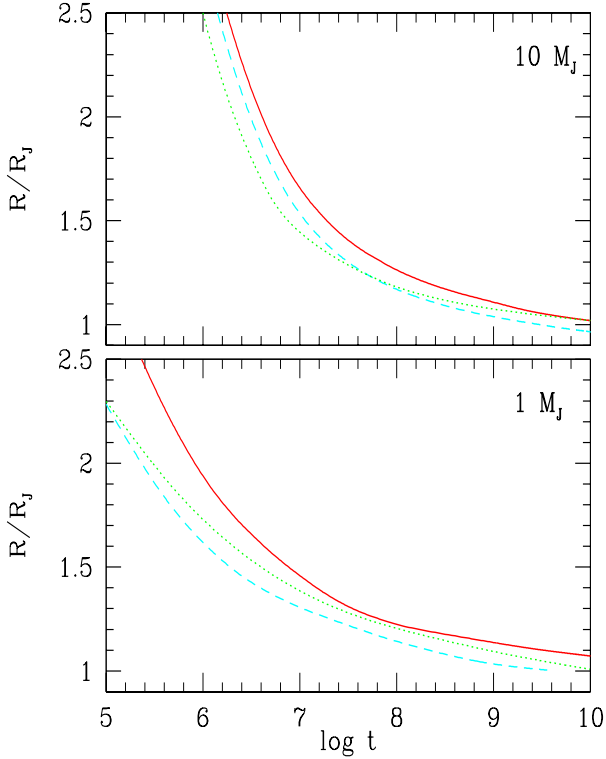,height=110mm,width=88mm} 
\caption{Radius (in $R_{\rm J}$) versus age (in yr) for EGP masses
of 1 $\mjup$ (lower panel) and 10 $\mjup$ (upper panel).
The dashed lines are nonirradiated models. The solid lines are
irradiated models 
at a separation $a$ = 0.046 AU from the parent star with $T_{{eff}_\star}$ = 6000K.
The dotted lines are nonirradiated models from Burrows et al. (1997).
}
\label{fig5}
\end{figure}

We have calculated the evolution of planets covering a range of mass from 0.5 $\mjup$
to 10 $\mjup$ with and without irradiation. The evolution of the radius as
a function of time is shown in Fig. \ref{fig5} for irradiated 
and nonirradiated
EGPs of 1 $\mjup$ and 10 $\mjup$. 
As expected, the less massive the planet, the larger
the effect of irradiation, for a given incident flux.
 At 1 Gyr, the 0.5 $\mjup$
EGP has a  14\% larger radius than its nonirradiated counterpart,
whereas for the 1 $\mjup$ (resp. 10  $\mjup$), $R$ is only 10\% larger 
(resp. 7\%).
We also compare our COND models (the nonirradiated models)
 to the Burrows et al. (1997, hereafter B97)
 nonirradiated models. Significant differences appear at young ages
 ($< 1$ Gyr), 
 due certainly to different initial conditions
(see Baraffe et al. 2002).
For ages $> 1$ Gyr and 
$m \simgr 5 \mjup$, the differences between the B97 models and ours 
 are of the order of the irradiation effects.
This reflects the different input physics, mainly
in the dust treatment and molecular opacities and illustrates the present uncertainties
in the models.
For $m \simle 5 \mjup$, however, irradiation effects become larger than
the differences between the B97 and our models.

\begin{figure}
\psfig{file=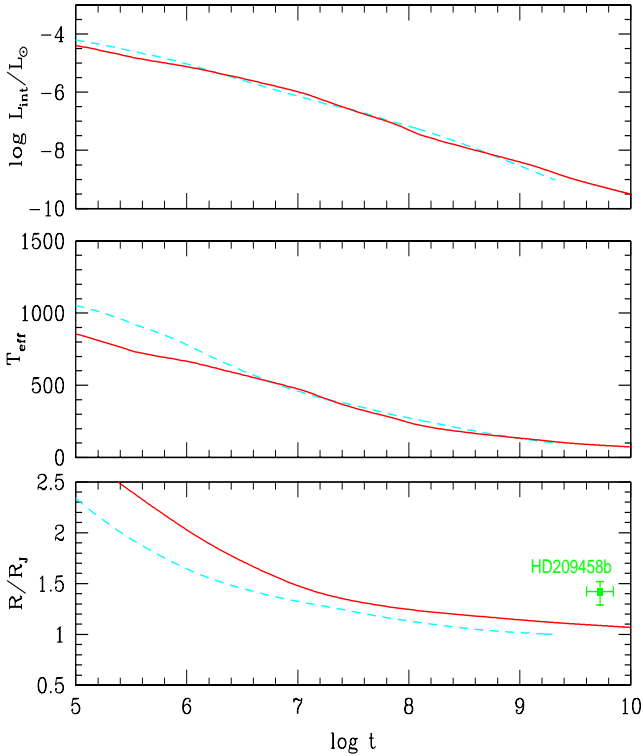,height=110mm,width=88mm} 
\caption{Effect of irradiation on the evolution of a planet with
$m_{\rm p}=0.69 \mjup$
at a separation $a$ = 0.046 AU from its parent star with $\te$ = 6000K.
The panels from top to bottom display respectively the intrinsic luminosity
$L_{\rm int}$, the effective temperature (in K) and the radius versus time (in yr). 
The solid curves correspond to the irradiated case and the dashed curves
 to the nonirradiated counterpart. 
We recall that in the case of irradiation,
$\te$ and $L_{\rm int}$ do not characterize
the total flux emitted by the planet. 
The position of HD209458b in the lower
panel is from Cody \& Sasselov (2002).
}
\label{fig6}
\end{figure}

The specific case of HD209458b, with a mass 
$m_{\rm p}=0.69 \mjup$, is illustrated
in Fig. \ref{fig6}. 
 The intrinsic luminosity and corresponding effective temperature
in the irradiated case (solid lines) are compared to the non-irradiated case (dashed lines).
Starting from the same initial configuration in both cases,
the heat loss is reduced at early ages in the case
of irradiation, as expected. Consequently, the irradiated model 
evolves at
larger entropy and radius than its non-irradiated counterpart. 
During the first Myr of evolution, both evolutionary sequences contract with increasing
central density and temperature, the non-irradiated model being
denser. The latter
becomes partially degenerate earlier, its contraction slows down and its heat loss becomes smaller than in the irradiated case (at $\log t \sim$ 6.2 yr). The situation reverses at $\log t \sim$ 7.4 yr when the effect of partial degeneracy becomes important in the irradiated sequence.
The age of  HD209458 is about  4-7 Gyr, according
to Cody \& Sasselov (2002). At 5 Gyr, the irradiated sequence
displayed in Fig. \ref{fig6} predicts a radius $R = 1.09 \rjup$, 26\%
smaller than the observed value. Without including irradiation effects,
the radius is $> $ 30\% than the observationally determined one.
Note that the nonirradiated
sequence stops at $\te$ = 100K, corresponding to an age of $\sim$ 2
Gyr and a radius $R = 1 \rjup$. In the following section, we analyse
the possible reasons for such a discrepancy.

\section{Discussion}

\subsection{Uncertainties of irradiated atmosphere/evolutionary models}

The question rises whether uncertainties of current models
can explain the mismatch of HD209458b predicted versus observed radius,
and whether irradiation effects can still provide the solution to the
problem.
We first note that our choice of parameters for the irradiated atmosphere
calculations certainly overestimates the effects of irradiation
(see \S 3). We  assume
redistribution of the incident flux 
over the day-side of the planet only, without
taking into account varying angles of incidence of the
impinging flux. As shown
in Brett \& Smith (1993), the effect of irradiation on the atmosphere
will decrease with increasing angles of incidence. 
Note that adopting the maximum case, i.e no redistribution
at all of $F_{\rm inc}$, affects significantly the outer
atmospheric profile, but only slightly the inner
profile.
We did not consider horizontal energy flows, which may exist
in a real planet receiving a varying amount of incident flux 
over its surface. But our assumption of isotropic incident flux,
with the maximum amount of flux allowed, should
provide an upper limit for the expected effects of irradiation
on the evolution of a planet.

Although present  atmosphere models
still have some shortcomings, due either to incomplete molecular opacities
or to dust treatment (see \S 2.2), the inner
profiles of the irradiated models 
must be changed and heated drastically to provide 
the effect required on evolutionary models to match the radius of
HD209458b.
Fig. \ref{fig6} indeed shows that the contraction of the irradiated sequence
proceeds too rapidly after the first Myr of evolution to maintain a radius
close to the observed value.
At an age $t_1$ = 1 Myr, the model has a radius
$R_1 = 2 R_{\rm J}$ and an intrinsic luminosity $\log L_1/L_\odot =
-5.12$, corresponding
to $\te$ = 670K.  This corresponds to
a characteristic Kelvin-Helmholtz
timescale $\tau_{\rm KH} = Gm_{\rm p}^2 /(R_1L_1) \sim 10$Myr.
In fact, the model has already reached a radius of $R_{10}= 1.48\,R_{\rm J}$,
about the radius of HD209458b, after only 10 Myr (see Fig. \ref{fig6}).
In order to slow down the planet contraction to reach the observed value at $\sim 5$ Gyr,
the inner profiles of the present irradiated
atmosphere models, for a given $\te$, should be drastically modified.
Such a major modification seems unlikely, for a fixed incident flux.
Note  that assuming a constant
flux from the parent star during the whole planet evolution
overestimates the total incident flux received by the planet
over $\sim$ 5 Gyr. 
For a parent star mass $\sim 1.06 \msol$ (Cody \& Sasselov 2002),
most of the  star evolution after
the first 1 Myr proceeds
at a luminosity $L$ smaller than its value at 5 Gyr. 

Finally, possible uncertainties due to abundance effects, such
as non-solar metallicity and/or helium abundance are not expected
to affect significantly the present results. To estimate 
such uncertainties,  
we have computed a grid of irradiated atmosphere models for 
an over-solar metallicity [M/H] = +0.3 and the corresponding 
evolutionary sequence for a mass $m_{\rm p}$ = 0.69 $\mjup$. After a few
Gyr evolution, the radius of the later sequence is essentially
the same as for the solar metallicity case.
Given the optimization of irradiation effects provided by
our assumptions, and the huge effect required on the inner profile
to reduce the mismatch between observed and predicted radii, we
do not expect uncertainties in the present models 
to be the source of the discrepancy.

\subsection{Observed versus theoretical radius}

The definition of the radius in low mass stars, brown dwarfs or
isolated giant planets is usually not a matter of confusion,
given the negligible extension of their atmosphere compared
to the total radius of the object. 
The picture could be different in the case of
irradiated atmospheres, where extension effects due to the large
heating of the upper layers may not be negligible (Seager \& Sasselov 2000;
Hubbard et al. 2001).

As discussed in Baschek et al. (1991), the condition of compactness for
a photosphere in  hydrostatic equilibrium is $H_{\rm P}/r << 1$, whit $H_{\rm P}$ the
pressure scale height and $r$ the radial distance to the center.
This condition is known to be perfectly fulfilled in 
(non-irradiated) low mass objects (see Chabrier \& Baraffe 1997),
where
the extension of the photosphere is usually less than 1\% the total 
size of the object. Thus the radius can unambiguously be defined, 
quoting Baschek et al. (1991), as the distance of the atmosphere 
to the object center.
In all our previous work, we fix the boundary condition between
atmosphere and inner structure  at $\tau_{\rm std}$=100, knowing that
$R(\tau_{\rm std} =100)$ is essentially the same as $R(\tau_{\rm
std} \sim 1)$ (see Chabrier \& Baraffe 1997). 
As already mentioned, we define $\tau_{\rm std}$ at 1.2 $\mu$m, 
which corresponds to the peak of the flux emitted by cool 
(non-irradiated) objects. Usually, 
the region where $\tau_{\rm std} \sim 1$ 
is close to the region where $\tau_{\rm Rosseland} \sim 1$ (for
the present irradiated models as well). 
Above this region, the atmosphere contains little mass and 
contributes negligibly to the luminosity.
The evolutionary calculations for irradiated models
presented in \S3.2 determine also the radius
at $\tau_{\rm std}=100$. Within the present assumptions of irradiation,
the atmospheric extension between $\tau_{\rm std}$=100 and
$\tau_{\rm
std} \sim 1$ represents only 1-2\% of the total radius. 
The presently calculated theoretical radius is thus essentially equivalent
to a photospheric radius at 1.2 $\mu$m.

However, the observed radius of HD209458b (e.g the one estimated
by Cody \& Sasselov 2002) is based on the analysis of optical
light curves.  It corresponds to a region of the atmosphere
where the optical depth is near unity at $0.6 \mu$m, which is
near $\tau_{\rm std} = 10^{-2}$  in our atmosphere models.  
Therefore, the radius predicted by
the evolutionary models is not equivalent to the observed
radius.  If the atmospheric extension  and the opacity of
the atmosphere are large at the observed wavelengths, the
measured radius could be different from the radius predicted
by the evolutionary calculation.  However, based on our irradiated atmosphere
models with the gravity predicted by the evolution and the age of
HD209458b (i.e.  $\log \, g$=3.2), the atmospheric extension between
$\tau_{\rm std} = 100$ and $\tau_{\rm std} = 10^{-2}$ (where the optical depth is 
close to unity at $0.6 \mu$ m) is very
small, namely 0.05 $\rjup$, compared to the overall radius (Barman et al. 2003, in preparation).  
Adding this value to the radius predicted by the models at 5 Gyr
yields an optical-depth radius at $0.6 \mu m$ of $\sim$ 1.14 $\rjup$, still 22\% less than the observed value. 
For younger planets or planets undergoing stronger irradiation
effects, with much lower gravities, the extension will be more
important and should be taken into account for a consistent comparison
between theoretical and observed radii, as already stressed
by Seager \& Sasselov (2000) and Hubbard et al. (2001).

In the same vein, Hubbard et al. (2001) estimate a radius of 94 430 km 
(1.32 $\rjup$) at a pressure of 1 bar, based on a detailed
analysis of physical effects influencing the observed light curve
of HD 209458b. At 5 Gyr, our models predict a radius at 1 bar
of $\sim$ 1.1 $\rjup$, 18\% smaller than the Hubbard et al. (2001) estimate.
Such a discrepancy is consistent with the afore-mentioned mismatch 
for the radius at $0.6 \mu$m.

\subsection{Other sources of energy deposit}

\begin{figure}
\psfig{file=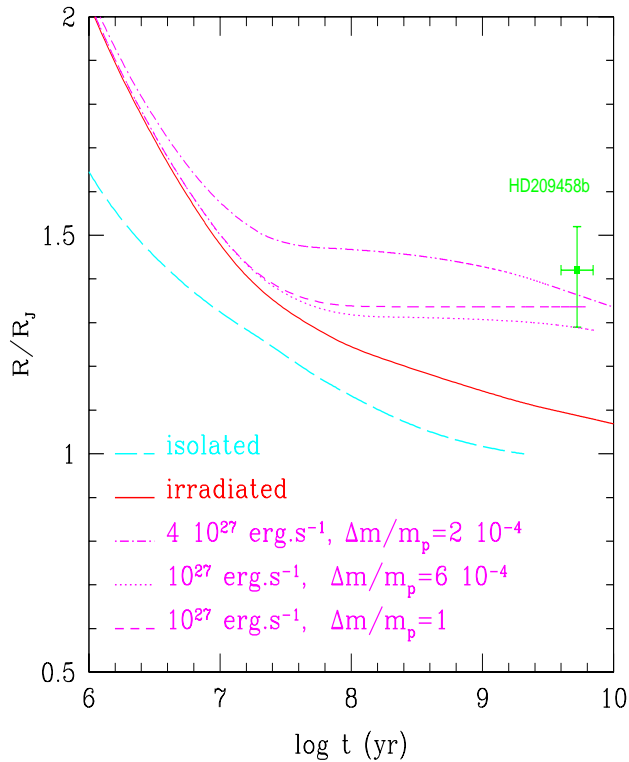,height=110mm,width=88mm} 
\caption{Effect of  extra source of energy dissipation on the
evolution of a planet with mass $m_{\rm p} = 0.69 \mjup$.
The solid  and long-dashed lines correspond to the
irradiated and non-irradiated sequences respectively, with
no extra source of energy dissipation. 
The other curves correspond to irradiated sequences with a total
amount of additional energy $L_{\rm extra}$ (in erg.s$^{-1}$)
deposited in layers between the surface and the mass
shell $m_{\rm p} - \Delta m$,
as indicated on  the figure (see text).
}
\label{fig7}
\end{figure}

If irradiation effects alone do not explain the large observed
radius of HD 209458b, other sources of energy must 
be invoked. Tidal interactions between the star and the planet
can provide a source of energy  associated
to the synchronization and/or circularization of the planet orbit,
dissipated within the planet (Lubow et al. 1997;
Rieutord \& Zahn 1997; Bodenheimer et al. 2001). However,
as discussed recently by Guillot \& Showman (2002) and Showman \&
Guillot (2002), these processes are efficient only during the early stages
of the planet evolution. 
Estimates based on the current understanding of such processes
yield  typical circularisation
timescale $\tau_{\rm circ} \sim 10^8$ yr (Bodenheimer et al. 2001)
and synchronisation timescale $\tau_{\rm syn} < 10^8$ yr (Lubow et
al. 1997; Rieutord \& Zahn 1997).
Such an energy source seems
thus unlikely to slow down the long term evolution of the planet,
 unless a second planet orbiting HD209458a is present. Such
a detection has been claimed very recently in the literature (Bodenheimer, Laughlin \& Lin, 2003) but remains to be confirmed unambiguously.
Showman \& Guillot (2002) suggested that downward transport
of kinetic energy produced by atmospheric circulation could be
dissipated in the planet interior, leading to a substantial deposit 
of energy. 
 Within the present input physics and treatment
of irradiation, we can estimate the amount of energy
required to reach the radius of HD 209458b. As in Guillot \& Showman (2002), we arbitrarily add an
extra term of energy generation $\dot \epsilon_{\rm extra}$ in the energy
equation at different depths. We have explored several cases
 displayed in Fig. \ref{fig7}. We add a total amount 
$L_{\rm extra} = \int \dot \epsilon_{\rm extra}dm$
 in a region  of mass $\Delta m$
enclosed between the surface and an arbitrary depth at mass shell
$m_1$
(i.e $\Delta m = m_{\rm p} -m_1$). Various tests indicate
that an amount of energy $L_{\rm extra} \sim 10^{27} -
5 \times10^{27}$ erg.s$^{-1}$ dissipated  along the internal adiabat 
 yields a radius
within the error bars of the observed value
(see Fig. \ref{fig7}). As expected, the larger 
the fraction of $L_{\rm extra}$ deposited in the
convective layers, the more important  the effect.  
Note that the case displayed in Fig. \ref{fig7}
with $L_{\rm extra} \sim 10^{27}$ erg.s$^{-1}$
dissipated all over the star ($\Delta m/m_{\rm p}$ = 1, dashed curve) is
equivalent to depositing the same amount of energy only at the
very center. Our quantitative estimates are in general agreement with 
Guillot \& Showman (2002). 
Such an amount of energy represents more than 100 times the intrinsic
luminosity $L_{int}$ of the planet, which is $\sim$ 10$^{25}$ erg.s$^{-1}$ at 1 Gyr
and $\sim$ 2 10$^{24}$ erg.s$^{-1}$ at 5 Gyr (see Fig. \ref{fig6}). However, it represents
only $\sim 1\%$ of the incoming luminosity, 
$L_{inc}=2\pi R_p^2 F_{inc}\sim 10^{29}$ erg s$^{-1}$, which largely dominates
the planet energetic balance.
Thus, an alternative possibility is the release of an external source of energy caused by
the incident radiation (see e.g. Showman \& Guillot, 2002).
As illustrated in Fig. \ref{fig7}, however, the extra source of energy must be dissipated at the top of
the internal adiabat, i.e. at a much deeper level than the penetration of the incident photons ($>>\tau=1$).

\section {Conclusion}

We have presented calculations describing the evolution
of cool brown dwarfs and extra-solar gaseous planets. The present
models reproduce the main trends of observed methane-dwarfs
in near-IR color-magnitude diagrams ($J$-$K$, $K$-$L$). Problems still remain
at wavelengths $<$ 1 $\mu$m, with a flux excess predicted
in the I-bandpass. The treatment of
atomic line broadening in such dense objects may be the source
of the present discrepancy. The models fail to reproduce
the coolest L-dwarfs and a detailed treatment of
dust diffusion/sedimentation is required for a correct description
of the transition region between L- and T- dwarfs. 
Work is in progress in this direction.

We have included the effects of irradiation,
{\it coupling  irradiated atmosphere profile and 
inner structure}, and providing
consistent evolutionary models for irradiated planets. 
The effect of irradiation are shown to modify significantly
the mechanical (mass-radius) and thermal evolution of
irradiated EGPs. 
However, a significant discrepancy 
(26\%) remains between the theoretical and observed radii of  
the transit planet HD209458b. 
We have explored possible uncertainties inherent in the
models to explain such a discrepancy.
Although solving these uncertainties
may modify the outer structure of the models
(extension of the atmosphere, albedo) and perhaps slightly reduce the discrepancy,
none of the uncertainties
is likely to modify significantly the inner entropy profile of the models,
which  determines the radius of the planet. 
Indeed, a drastic modification
on the inner thermal structure is required to bring the theoretical
radius in agreement with the observed one.


In summary, we do not expect irradiation effects alone to explain the
large observed radius of HD209458b.  In the same vein, tidal interactions
will affect only the early stages of evolution of the planet but will probably be
dissipated too rapidly to affect the long term contraction of the object. Other sources of energy,
representing about 100 times the intrinsic luminosity of the planet, 
seem to be required to explain the observed radius.
The first extra-solar planet transit thus remains a challenge for theory.
Detection of other transits is now crucial to conclude whether 
HD209458 is a peculiar system, whether a second planetary companion
is confirmed or not,  or whether we are missing something
in the current understanding of close-in giant planets.

\medskip
{\it Note: }
Isochrones for $t \, \ge$ 1 Myr of the COND models
(from 0.5 $\mjup$ to 0.1 $\msol$) are available at:
\\
\hskip 1cm http://www.ens-lyon.fr/\~\,ibaraffe/COND03\_models

\begin{acknowledgements} 
We are very grateful to H. Harris and S. Leggett for providing 
data under ascii files and to Doug Lin for mentioning the possible
detection of a second planet. We thank our anonymous referee for valuable comments.
I.B thanks the Max-Planck Institut f\"ur Astrophysik in Garching
for hospitality during elaboration of part of this work.
This research was supported in part by the LTSA grant NAG 5-3435,
the NASA EPSCor grant to Wichita State University,
NSF grants AST-9720704 and  
AST-0086246, NASA grants NAG5-8425, NAG5-9222, as well as NASA/JPL  
grant 961582 to the University of Georgia. This work was supported in  
part by the P\^ole
Scientifique de Mod\'elisation Num\'erique at ENS-Lyon.  Some of the
calculations presented in this paper were performed on 
the IBM pSeries  690 of
the Norddeutscher Verbund f\"ur Hoch- und H\"ochstleistungsrechnen  
(HLRN), on
the IBM SP ``Blue Horizon'' of the San Diego Supercomputer Center  
(SDSC), with
support from the National Science Foundation, on the IBM SP 
and the Cray
T3E of the NERSC with support from the DoE,
and using the computer facilities at Centre
d'Etudes Nucl\'eaires de Grenoble, CINES and  IDRIS. 
We thank all these  
institutions for a generous allocation of computer time.
\end{acknowledgements}

\begin{table*}
\caption{COND isochrones for 0.1 Gyr
}
\begin{tabular}{lcccccccccccc}
\hline\noalign{\smallskip}
$m/\msol$  &$T_{eff}$ & $\log L/\lsol$ & $R/R_\odot$ & $\log \,g$
&$M_V$ &$M_R$ &$M_I$ &$M_J$  &$M_H$&
 $M_K$ & $M_{L'}$ & $M_M$\\
\noalign{\smallskip}
\hline\noalign{\smallskip}
 0.0005&  240. & -7.418& 0.114& 3.020& 41.98& 37.51& 34.00& 28.42& 26.59& 37.66& 19.57& 17.64\\
 0.0010&  309. & -6.957& 0.117& 3.300& 32.58& 28.68& 25.89& 22.43& 22.38& 29.11& 17.41& 15.69\\
 0.0020&  425. & -6.383& 0.120& 3.580& 29.69& 25.62& 22.79& 20.05& 19.76& 23.13& 15.94& 14.55\\
 0.0030&  493. & -6.112& 0.121& 3.746& 28.71& 24.48& 21.66& 18.88& 18.57& 20.88& 15.21& 13.93\\
 0.0040&  563. & -5.880& 0.122& 3.869& 28.09& 23.77& 20.95& 17.95& 17.71& 19.35& 14.59& 13.50\\
 0.0050&  630. & -5.686& 0.122& 3.965& 27.65& 23.25& 20.44& 17.23& 17.02& 18.15& 14.06& 13.14\\
 0.0060&  688. & -5.534& 0.121& 4.048& 27.36& 22.92& 20.09& 16.71& 16.51& 17.26& 13.67& 12.83\\
 0.0070&  760. & -5.365& 0.121& 4.117& 27.03& 22.55& 19.74& 16.16& 16.01& 16.38& 13.26& 12.55\\
 0.0080&  816. & -5.246& 0.120& 4.180& 26.77& 22.28& 19.49& 15.76& 15.65& 15.79& 12.97& 12.35\\
 0.0090&  886. & -5.103& 0.120& 4.232& 26.45& 21.96& 19.19& 15.32& 15.23& 15.16& 12.63& 12.13\\
 0.0100&  953. & -4.978& 0.120& 4.279& 26.10& 21.66& 18.92& 14.94& 14.86& 14.69& 12.34& 11.96\\
 0.0120& 1335. & -4.332& 0.129& 4.297& 23.53& 19.44& 16.79& 13.20& 12.97& 12.76& 10.90& 11.17\\
 0.0150& 1399. & -4.281& 0.124& 4.424& 23.30& 19.24& 16.46& 13.05& 12.82& 12.65& 10.83& 11.15\\
 0.0200& 1561. & -4.110& 0.122& 4.569& 22.30& 18.55& 16.08& 12.60& 12.34& 12.17& 10.53& 10.99\\
 0.0300& 1979. & -3.668& 0.126& 4.715& 19.96& 16.80& 14.48& 11.52& 11.20& 10.90&  9.82& 10.38\\
 0.0400& 2270. & -3.386& 0.132& 4.797& 18.46& 15.63& 13.31& 10.89& 10.52& 10.19&  9.39&  9.84\\
 0.0500& 2493. & -3.167& 0.141& 4.837& 17.09& 14.77& 12.53& 10.43& 10.02&  9.71&  9.04&  9.37\\
 0.0600& 2648. & -3.008& 0.150& 4.863& 16.08& 14.12& 12.01& 10.10&  9.68&  9.37&  8.78&  9.03\\
 0.0700& 2762. & -2.879& 0.160& 4.874& 15.33& 13.59& 11.60&  9.82&  9.39&  9.10&  8.55&  8.75\\
 0.0720& 2782. & -2.856& 0.162& 4.875& 15.20& 13.50& 11.53&  9.77&  9.34&  9.05&  8.51&  8.70\\
 0.0750& 2809. & -2.821& 0.166& 4.875& 15.01& 13.36& 11.42&  9.69&  9.26&  8.97&  8.44&  8.63\\
 0.0800& 2846. & -2.776& 0.170& 4.880& 14.77& 13.18& 11.29&  9.60&  9.16&  8.87&  8.36&  8.53\\
 0.0900& 2910. & -2.689& 0.180& 4.884& 14.34& 12.85& 11.03&  9.40&  8.96&  8.68&  8.19&  8.35\\
 0.1000& 2960. & -2.617& 0.189& 4.887& 14.02& 12.58& 10.82&  9.24&  8.80&  8.52&  8.05&  8.19\\

\hline
\end{tabular}
\begin{list}{}
\item Notes.--
$T_{eff}$ is in K, the gravity $g$ in cgs.
 The VRI magnitudes are in the
Johnson-Cousins system (Bessell 1990), JHK in the CIT system (Leggett
1992), L$^\prime$ in the Johnson-Glass system and M in the Johnson system.
\end{list}
\end{table*}

\begin{table*}
\caption{Same as Table 1 for 0.5 Gyr
}
\begin{tabular}{lcccccccccccc}
\hline\noalign{\smallskip}
$m/\msol$  &$T_{eff}$ & $\log L/\lsol$ & $R/R_\odot$ & $\log \,g$
&$M_V$ &$M_R$ &$M_I$ &$M_J$  &$M_H$&
 $M_K$ & $M_{L'}$ & $M_M$\\
\noalign{\smallskip}
\hline\noalign{\smallskip}
 0.0005&  141. & -8.415& 0.105& 3.097& 56.30& 51.03& 46.60& 37.42& 33.07& 51.62& 23.09& 20.59\\
 0.0010&  203. & -7.753& 0.109& 3.365& 47.57& 42.88& 38.99& 31.61& 29.15& 43.23& 20.93& 18.68\\
 0.0020&  272. & -7.218& 0.112& 3.639& 37.05& 33.00& 30.06& 25.07& 24.62& 34.02& 18.66& 16.58\\
 0.0030&  322. & -6.913& 0.113& 3.805& 32.02& 28.23& 25.75& 22.05& 22.27& 29.03& 17.52& 15.53\\
 0.0040&  370. & -6.670& 0.114& 3.928& 30.65& 26.73& 24.16& 21.01& 21.06& 26.45& 16.85& 15.05\\
 0.0050&  409. & -6.496& 0.113& 4.027& 29.60& 25.57& 22.94& 20.20& 20.11& 24.54& 16.32& 14.67\\
 0.0060&  449. & -6.340& 0.113& 4.112& 29.16& 25.05& 22.39& 19.64& 19.51& 23.10& 15.92& 14.36\\
 0.0070&  488. & -6.200& 0.112& 4.185& 28.71& 24.51& 21.80& 19.10& 18.91& 21.68& 15.52& 14.06\\
 0.0080&  525. & -6.080& 0.111& 4.249& 28.40& 24.14& 21.41& 18.65& 18.46& 20.74& 15.19& 13.83\\
 0.0090&  564. & -5.963& 0.110& 4.307& 28.14& 23.82& 21.07& 18.21& 18.04& 19.95& 14.88& 13.63\\
 0.0100&  599. & -5.864& 0.110& 4.358& 27.91& 23.53& 20.77& 17.80& 17.66& 19.22& 14.59& 13.43\\
 0.0120&  759. & -5.447& 0.110& 4.432& 27.20& 22.73& 19.91& 16.38& 16.29& 16.82& 13.51& 12.68\\
 0.0150&  791. & -5.404& 0.107& 4.557& 27.11& 22.63& 19.81& 16.20& 16.16& 16.58& 13.41& 12.60\\
 0.0200&  936. & -5.133& 0.104& 4.704& 26.53& 22.07& 19.27& 15.33& 15.34& 15.37& 12.78& 12.21\\
 0.0300& 1264. & -4.636& 0.101& 4.905& 24.97& 20.77& 17.95& 13.90& 13.87& 13.69& 11.70& 11.68\\
 0.0400& 1583. & -4.255& 0.100& 5.040& 23.11& 19.28& 16.74& 12.92& 12.76& 12.68& 10.95& 11.33\\
 0.0500& 1875. & -3.955& 0.101& 5.131& 21.31& 17.94& 15.53& 12.21& 11.95& 11.79& 10.44& 10.99\\
 0.0600& 2116. & -3.729& 0.102& 5.194& 19.99& 16.91& 14.50& 11.69& 11.36& 11.13& 10.10& 10.63\\
 0.0700& 2329. & -3.534& 0.106& 5.233& 18.84& 16.04& 13.68& 11.27& 10.90& 10.63&  9.81& 10.26\\
 0.0720& 2369. & -3.498& 0.107& 5.238& 18.60& 15.89& 13.54& 11.19& 10.81& 10.54&  9.75& 10.19\\
 0.0750& 2426. & -3.445& 0.108& 5.244& 18.25& 15.66& 13.33& 11.08& 10.69& 10.42&  9.67& 10.08\\
 0.0800& 2518. & -3.356& 0.111& 5.248& 17.65& 15.27& 13.00& 10.89& 10.49& 10.22&  9.53&  9.89\\
 0.0900& 2680. & -3.189& 0.119& 5.241& 16.54& 14.56& 12.43& 10.55& 10.12&  9.85&  9.25&  9.53\\
 0.1000& 2804. & -3.047& 0.128& 5.223& 15.68& 13.96& 11.98& 10.25&  9.80&  9.54&  9.00&  9.22\\

\hline
\end{tabular}
\end{table*}

\begin{table*}
\caption{Same as Table 1 for 1 Gyr
}
\begin{tabular}{lcccccccccccc}
\hline\noalign{\smallskip}
$m/\msol$  &$T_{eff}$ & $\log L/\lsol$ & $R/R_\odot$ & $\log \,g$
&$M_V$ &$M_R$ &$M_I$ &$M_J$  &$M_H$&
 $M_K$ & $M_{L'}$ & $M_M$\\
\noalign{\smallskip}
\hline\noalign{\smallskip}
 0.0005&  111. & -8.851& 0.102& 3.115& 60.75& 55.23& 50.50& 40.19& 35.07& 55.87& 24.15& 21.49\\
 0.0010&  160. & -8.185& 0.106& 3.386& 54.15& 49.10& 44.69& 35.58& 32.06& 49.17& 22.40& 19.95\\
 0.0020&  226. & -7.560& 0.109& 3.662& 44.39& 39.91& 36.34& 29.28& 27.80& 40.31& 20.18& 17.94\\
 0.0030&  270. & -7.244& 0.111& 3.827& 37.64& 33.60& 30.73& 25.29& 24.99& 34.51& 18.84& 16.66\\
 0.0040&  304. & -7.031& 0.111& 3.950& 32.62& 28.93& 26.60& 22.49& 22.91& 30.25& 17.91& 15.73\\
 0.0050&  342. & -6.831& 0.110& 4.051& 31.58& 27.79& 25.36& 21.71& 21.98& 28.28& 17.38& 15.36\\
 0.0060&  377. & -6.664& 0.110& 4.134& 30.53& 26.63& 24.12& 20.96& 21.07& 26.39& 16.87& 15.01\\
 0.0070&  403. & -6.556& 0.109& 4.208& 29.77& 25.79& 23.26& 20.41& 20.43& 25.04& 16.54& 14.76\\
 0.0080&  438. & -6.417& 0.108& 4.272& 29.37& 25.31& 22.73& 19.93& 19.89& 23.77& 16.18& 14.50\\
 0.0090&  464. & -6.325& 0.107& 4.331& 29.06& 24.94& 22.33& 19.59& 19.49& 22.85& 15.93& 14.31\\
 0.0100&  491. & -6.235& 0.107& 4.383& 28.74& 24.55& 21.89& 19.23& 19.07& 21.90& 15.65& 14.11\\
 0.0120&  578. & -5.955& 0.106& 4.467& 28.09& 23.75& 21.01& 18.15& 18.03& 19.86& 14.88& 13.60\\
 0.0150&  628. & -5.835& 0.103& 4.587& 27.86& 23.47& 20.72& 17.70& 17.62& 19.05& 14.56& 13.38\\
 0.0200&  766. & -5.514& 0.100& 4.736& 27.31& 22.85& 20.05& 16.56& 16.55& 17.19& 13.72& 12.80\\
 0.0300& 1009. & -5.071& 0.096& 4.948& 26.40& 21.96& 19.15& 15.10& 15.16& 15.14& 12.67& 12.15\\
 0.0400& 1271. & -4.696& 0.093& 5.099& 25.19& 20.99& 18.13& 14.04& 14.04& 13.90& 11.88& 11.80\\
 0.0500& 1543. & -4.374& 0.092& 5.211& 23.73& 19.81& 17.15& 13.21& 13.12& 13.04& 11.25& 11.53\\
 0.0600& 1801. & -4.106& 0.092& 5.291& 22.13& 18.59& 16.10& 12.56& 12.36& 12.27& 10.77& 11.26\\
 0.0700& 2082. & -3.829& 0.094& 5.333& 20.44& 17.31& 14.87& 11.93& 11.62& 11.43& 10.32& 10.85\\
 0.0720& 2140. & -3.772& 0.095& 5.336& 20.12& 17.05& 14.61& 11.80& 11.48& 11.27& 10.23& 10.75\\
 0.0750& 2234. & -3.679& 0.098& 5.334& 19.59& 16.63& 14.21& 11.60& 11.25& 11.02& 10.08& 10.58\\
 0.0800& 2383. & -3.527& 0.102& 5.323& 18.67& 15.96& 13.61& 11.26& 10.89& 10.63&  9.84& 10.27\\
 0.0900& 2627. & -3.268& 0.113& 5.285& 16.98& 14.86& 12.67& 10.72& 10.30& 10.03&  9.40&  9.72\\
 0.1000& 2784. & -3.083& 0.125& 5.246& 15.86& 14.09& 12.09& 10.33&  9.89&  9.62&  9.07&  9.31\\

\hline
\end{tabular}
\end{table*}

\begin{table*}
\caption{Same as Table 1 for 5 Gyr
}
\begin{tabular}{lcccccccccccc}
\hline\noalign{\smallskip}
$m/\msol$  &$T_{eff}$ & $\log L/\lsol$ & $R/R_\odot$ & $\log \,g$
&$M_V$ &$M_R$ &$M_I$ &$M_J$  &$M_H$&
 $M_K$ & $M_{L'}$ & $M_M$\\
\noalign{\smallskip}
\hline\noalign{\smallskip}
 0.0020&  129. & -8.570& 0.105& 3.698& 60.05& 54.63& 49.68& 38.16& 34.52& 53.62& 23.33& 20.80\\
 0.0030&  162. & -8.166& 0.105& 3.868& 55.32& 50.15& 45.63& 34.96& 32.46& 49.21& 22.22& 19.81\\
 0.0040&  193. & -7.867& 0.105& 3.994& 50.68& 45.80& 41.76& 32.14& 30.52& 45.15& 21.28& 18.92\\
 0.0050&  220. & -7.644& 0.105& 4.095& 46.50& 41.92& 38.30& 29.83& 28.84& 41.71& 20.48& 18.16\\
 0.0060&  244. & -7.469& 0.104& 4.180& 42.71& 38.43& 35.18& 27.80& 27.34& 38.67& 19.78& 17.49\\
 0.0070&  265. & -7.328& 0.103& 4.254& 39.29& 35.29& 32.37& 26.02& 26.00& 35.96& 19.16& 16.88\\
 0.0080&  284. & -7.217& 0.103& 4.318& 36.31& 32.57& 29.93& 24.50& 24.85& 33.65& 18.64& 16.36\\
 0.0090&  301. & -7.124& 0.102& 4.376& 33.73& 30.22& 27.82& 23.21& 23.85& 31.64& 18.18& 15.91\\
 0.0100&  322. & -7.015& 0.101& 4.429& 33.05& 29.47& 27.06& 22.74& 23.30& 30.49& 17.91& 15.71\\
 0.0120&  361. & -6.823& 0.100& 4.519& 31.58& 27.86& 25.46& 21.77& 22.17& 28.16& 17.41& 15.32\\
 0.0150&  399. & -6.671& 0.098& 4.634& 30.20& 26.35& 24.00& 20.85& 21.14& 26.01& 16.96& 14.97\\
 0.0200&  473. & -6.401& 0.095& 4.786& 29.24& 25.17& 22.66& 19.89& 19.92& 23.31& 16.18& 14.45\\
 0.0300&  610. & -6.008& 0.090& 5.011& 28.17& 23.83& 21.17& 18.34& 18.31& 20.04& 15.04& 13.71\\
 0.0400&  760. & -5.670& 0.085& 5.179& 27.58& 23.15& 20.40& 17.04& 17.11& 17.99& 14.17& 13.13\\
 0.0500&  931. & -5.353& 0.082& 5.313& 27.09& 22.63& 19.82& 15.94& 16.05& 16.38& 13.38& 12.63\\
 0.0600& 1120. & -5.058& 0.079& 5.418& 26.44& 22.03& 19.20& 15.01& 15.13& 15.15& 12.73& 12.27\\
 0.0700& 1524. & -4.504& 0.081& 5.466& 24.33& 20.36& 17.60& 13.52& 13.50& 13.44& 11.60& 11.77\\
 0.0720& 1712. & -4.278& 0.083& 5.453& 23.11& 19.36& 16.75& 12.97& 12.85& 12.80& 11.16& 11.54\\
 0.0750& 2006. & -3.942& 0.089& 5.411& 21.03& 17.80& 15.34& 12.20& 11.92& 11.78& 10.54& 11.06\\
 0.0800& 2320. & -3.603& 0.099& 5.353& 19.11& 16.28& 13.89& 11.44& 11.07& 10.82&  9.97& 10.43\\
 0.0900& 2622. & -3.275& 0.113& 5.289& 17.02& 14.88& 12.70& 10.73& 10.31& 10.04&  9.42&  9.73\\
 0.1000& 2785. & -3.083& 0.125& 5.247& 15.85& 14.09& 12.09& 10.33&  9.89&  9.62&  9.07&  9.31\\
\hline
\end{tabular}
\end{table*}

\begin{table*}
\caption{Same as Table 1 for 10 Gyr
}
\begin{tabular}{lcccccccccccc}
\hline\noalign{\smallskip}
$m/\msol$  &$T_{eff}$ & $\log L/\lsol$ & $R/R_\odot$ & $\log \,g$
&$M_V$ &$M_R$ &$M_I$ &$M_J$  &$M_H$&
 $M_K$ & $M_{L'}$ & $M_M$\\
\noalign{\smallskip}
\hline\noalign{\smallskip}
 0.0030&  125. & -8.629& 0.104& 3.879& 61.46& 55.89& 50.79& 38.29& 35.04& 54.28& 23.36& 20.89\\
 0.0040&  149. & -8.325& 0.104& 4.006& 58.00& 52.62& 47.86& 36.01& 33.56& 51.11& 22.59& 20.19\\
 0.0050&  172. & -8.087& 0.103& 4.109& 54.62& 49.46& 45.05& 34.06& 32.18& 48.26& 21.92& 19.57\\
 0.0060&  193. & -7.888& 0.102& 4.195& 51.29& 46.38& 42.30& 32.23& 30.85& 45.53& 21.29& 18.97\\
 0.0070&  213. & -7.724& 0.102& 4.270& 48.17& 43.50& 39.72& 30.55& 29.62& 43.01& 20.71& 18.41\\
 0.0080&  232. & -7.584& 0.101& 4.335& 45.19& 40.77& 37.28& 29.00& 28.46& 40.66& 20.18& 17.89\\
 0.0090&  249. & -7.469& 0.100& 4.393& 42.49& 38.30& 35.07& 27.62& 27.41& 38.55& 19.70& 17.42\\
 0.0100&  265. & -7.368& 0.099& 4.445& 39.95& 35.99& 33.00& 26.34& 26.44& 36.59& 19.26& 16.98\\
 0.0120&  293. & -7.204& 0.098& 4.536& 35.23& 31.68& 29.12& 24.04& 24.64& 33.06& 18.48& 16.20\\
 0.0150&  330. & -7.016& 0.096& 4.650& 32.81& 29.25& 26.80& 22.70& 23.33& 30.46& 17.93& 15.76\\
 0.0200&  389. & -6.759& 0.093& 4.802& 30.75& 26.99& 24.64& 21.28& 21.73& 27.02& 17.21& 15.19\\
 0.0300&  504. & -6.358& 0.088& 5.029& 28.98& 24.87& 22.35& 19.72& 19.68& 22.59& 16.04& 14.37\\
 0.0400&  634. & -6.004& 0.083& 5.200& 28.18& 23.85& 21.19& 18.33& 18.33& 19.99& 15.03& 13.75\\
 0.0500&  776. & -5.695& 0.079& 5.338& 27.64& 23.21& 20.47& 17.10& 17.18& 18.07& 14.23& 13.21\\
 0.0600&  941. & -5.393& 0.076& 5.450& 27.20& 22.74& 19.93& 16.04& 16.18& 16.55& 13.50& 12.73\\
 0.0700& 1289. & -4.832& 0.078& 5.503& 25.69& 21.45& 18.60& 14.37& 14.43& 14.36& 12.27& 12.08\\
 0.0720& 1556. & -4.472& 0.081& 5.481& 24.17& 20.22& 17.48& 13.44& 13.41& 13.36& 11.55& 11.75\\
 0.0750& 1997. & -3.954& 0.089& 5.415& 21.10& 17.85& 15.39& 12.23& 11.95& 11.81& 10.56& 11.08\\
 0.0800& 2322. & -3.602& 0.099& 5.353& 19.10& 16.27& 13.89& 11.43& 11.06& 10.82&  9.97& 10.43\\
 0.0900& 2624. & -3.274& 0.113& 5.289& 17.01& 14.88& 12.69& 10.73& 10.31& 10.04&  9.41&  9.73\\
 0.1000& 2786. & -3.082& 0.125& 5.246& 15.85& 14.09& 12.08& 10.32&  9.88&  9.62&  9.07&  9.30\\
\hline
\end{tabular}
\end{table*}

\end{document}